\documentclass[aip,amsmath,amssymb,reprint]{revtex4-1}

\usepackage{graphicx}
\usepackage{dcolumn}
\usepackage{bm}
\usepackage[utf8]{inputenc}
\usepackage[T1]{fontenc}
\usepackage{mathptmx}

\begin{document}

\title{Nanoassembly technique of carbon nanotubes for hybrid circuit-QED} 

\author{T. Cubaynes}
\affiliation{$^{1}$Laboratoire de Physique de l'Ecole normale sup\'{e}rieure, ENS,
Universit\'{e} PSL, CNRS, Sorbonne
Universit\'{e}, Universit\'{e} Paris-Diderot, Sorbonne Paris Cit\'{e},
Paris, France}

\author{L.C. Contamin}
\affiliation{$^{1}$Laboratoire de Physique de l'Ecole normale sup\'{e}rieure, ENS,
Universit\'{e} PSL, CNRS, Sorbonne
Universit\'{e}, Universit\'{e} Paris-Diderot, Sorbonne Paris Cit\'{e},
Paris, France}

\author{M.C. Dartiailh}
\affiliation{$^{1}$Laboratoire de Physique de l'Ecole normale sup\'{e}rieure, ENS,
Universit\'{e} PSL, CNRS, Sorbonne
Universit\'{e}, Universit\'{e} Paris-Diderot, Sorbonne Paris Cit\'{e},
Paris, France}

\author{M.M. Desjardins}
\affiliation{$^{1}$Laboratoire de Physique de l'Ecole normale sup\'{e}rieure, ENS,
Universit\'{e} PSL, CNRS, Sorbonne
Universit\'{e}, Universit\'{e} Paris-Diderot, Sorbonne Paris Cit\'{e},
Paris, France}

\author{A. Cottet}
\affiliation{$^{1}$Laboratoire de Physique de l'Ecole normale sup\'{e}rieure, ENS,
Universit\'{e} PSL, CNRS, Sorbonne
Universit\'{e}, Universit\'{e} Paris-Diderot, Sorbonne Paris Cit\'{e},
Paris, France}

\author{M.R. Delbecq}
\affiliation{$^{1}$Laboratoire de Physique de l'Ecole normale sup\'{e}rieure, ENS,
Universit\'{e} PSL, CNRS, Sorbonne
Universit\'{e}, Universit\'{e} Paris-Diderot, Sorbonne Paris Cit\'{e},
Paris, France}

\author{T. Kontos}
\affiliation{$^{1}$Laboratoire de Physique de l'Ecole normale sup\'{e}rieure, ENS,
Universit\'{e} PSL, CNRS, Sorbonne
Universit\'{e}, Universit\'{e} Paris-Diderot, Sorbonne Paris Cit\'{e},
Paris, France}

\date{\today}

\begin{abstract}

A complex quantum dot circuit based on a clean and suspended carbon nanotube embedded in a circuit quantum electrodynamique (cQED) architecture is a very attractive platform to investigate a large spectrum of physics phenomena ranging from qubit physics to nanomechanics. We demonstrate a carbon nanotube transfer process allowing us to integrate clean carbon nanotubes into complex quantum dot circuits inside a cQED platform. This technique is compatible with various contacting materials such as superconductors or ferromagnets. This makes it suitable for hybrid quantum devices. Our results are based on 8 different devices demonstrating the robustness of this technique.
\end{abstract}

\pacs{}

\maketitle 

One challenge for the investigation of quantum phenomena is the fabrication of systems which are sufficiently decoupled from their environment such that they manifest their quantum nature. This decoupling usually comes at the cost of low tunability of the system parameters. In this perspective, a suspended and clean carbon nanotube embedded in a circuit, hence allowing large tunability of its parameters, and with the “active” part of the system confined to the suspended section of the nanotube, is very attractive \cite{Waissman2013c}.

Because it can be suspended, a carbon nanotube can be engineered to be arbitrarily far from any interface which is known to be one of the main source of charge noise \cite{Sharf2012}. The cleanliness of the nanotube plays also a key role since it ensures a very low amount of charge fluctuators. These two points suggest long electronic coherence time in such carbon nanotube based circuits. Beside, there is also the possibility to taylor the phonon spectrum of the suspended nanotube via gate voltages \cite{Sazonova2004, Witkamp2006, Lassagne2008}. In case of a qubit, this means that one can tune the relaxation time due to electron-phonon interaction \citep{Cottet2010}. The nanotube can also be connected with different types of metals, making this platform interesting for studying hybrid circuit \cite{Cottet2017}.

Recently, various carbon nanotube transfer techniques have been developped \cite{Wu2010a, Muoth2012, Pei2012, Waissman2013c, Ranjan2015, Blien2018a, Blien2020}. The common idea is to grow carbon nanotubes on a separate substrate, then transfering one of the nanotubes onto the circuit at the final step of the process. Depending on the growth and the transfer conditions, several of these works have demonstrated the clean nature of the transfered nanotube \cite{Pei2012, Waissman2013c, Ranjan2015}, up to the observation of a 1D Wigner crystal \cite{Shapir2019}.

This technique has the advantage to give suspended carbon nanotubes, and is now largely used in nanomechanics experiments \cite{Benyamini2014, Khivrich2019a, Wen2019b}. The suspended nature of the nanotube is also desirable to limit charge noise in quantum dot circuits. Also, because the fabrication of the circuit and the synthesis of the carbon nanotubes are now completely independent, there is a great flexibility in the circuit fabrication process. More recently this technique has been adapted to cQED platform \cite{Ranjan2015, Blien2020}, by growing carbon nanotubes on a fork-like chip \cite{Muoth2012, Blien2018a}.

Here we present a fabrication technique inheriting from these recent advances. The nanotubes are grown on a substrate containing a comb of 48 cantilevers allowing to do a pre-selection of the transferred nanotube. In addition to producing clean suspended carbon nanotube based circuits embedded in a cQED platform, we show that this technique is compatible with superconductor and ferromagnet as metallic contacts, making it suitable for many types of hybrid quantum devices. Our results are based on 8 devices with different geometry and contact materials. 

\begin{figure}[b]
\includegraphics[width=0.95\linewidth]{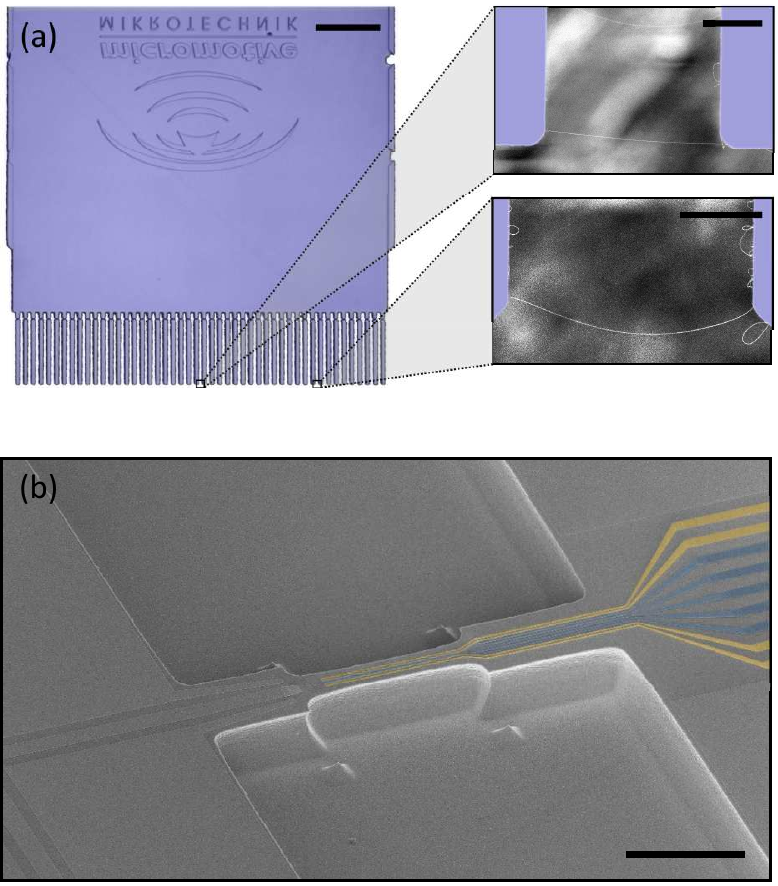}
\caption{\label{figure1}
\textbf{a.} Optical photograph of the growth chip. It is composed of 48 cantilevers which are visible on the bottom edge of the chip. scale bar: 500 $\mu $m. The zooms in on the cantilevers show isolaled carbon nanotubes suspended in between two adjacent cantilevers. Scale bar: 10 $\mu$m.
\textbf{b.} False color scanning electron micrograph of the circuit chip (Device 5). The circuit, constituted of the gate electrodes (in dark blue) and the contact electrode (in yellow), is sandwiched in between the two trenches. Scale bar: 50 $\mu$m. }
\end{figure}

\begin{figure*}[t!]
\includegraphics[width=0.95\linewidth]{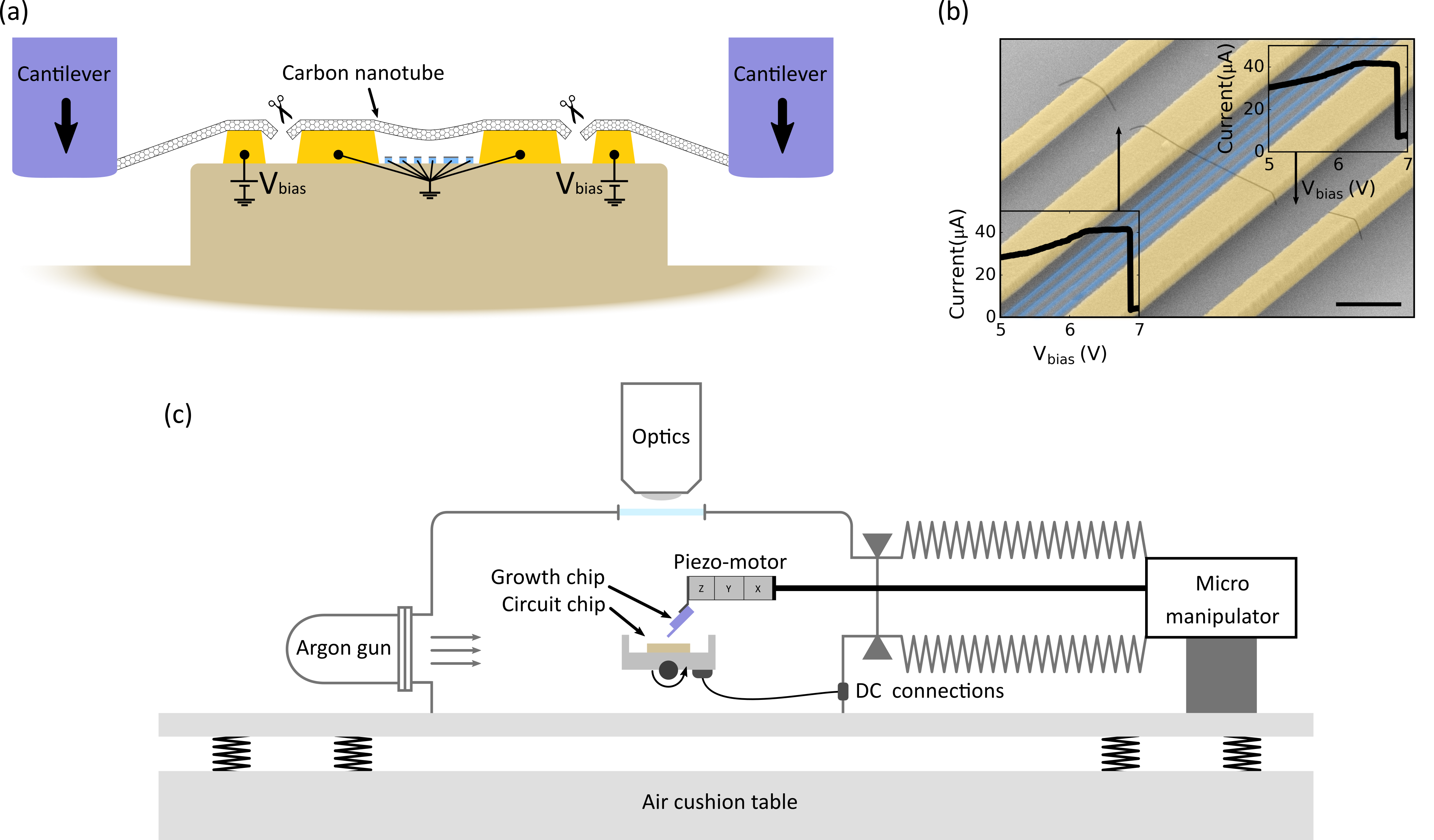}
\caption{\label{figure2}
Principle of the transfer technique.
\textbf{a.} Schematics of the carbon nanotube transfer process. The cantilevers (in purple) are lowered until the carbon nanotube makes the connection between external and internal contacts (in orange) of the circuit chip. The nanotube is then cut on both side of the circuit by successively biasing only one of the two external contacts.
\textbf{b.} False color scanning electron micrograph of the region where the nanotube is integrated (Device 1). The carbon nanotube (in black) is suspended over a gate array (in dark blue). Scale bar: 1$\mu$m. Insets: Current as a function of the bias voltage applied on the left external contact (bottom-left inset) and on the right external contact (top-right inset). 
\textbf{c.} Schematic of the vacuum chamber used for the transfer of the carbon nanotube. The argon gun on the left side of the chamber is used for cleaning the circuit chip, or to remove unwanted carbon nanotubes. The growth chip (in purple) can be locked-up in the bellows chamber on the right side of the chamber, for protecting the nanotubes during ventilation of the main chamber or the use of the argon gun.
}
\end{figure*}

The process includes two distinct samples: the growth chip, on which carbon nanotubes are grown, and the circuit chip. For the growth chip, we use a commercial cantilever chip with 48 cantilevers and a spacing of 30 $\mu$m between cantilevers as depicted in figure \ref{figure1}.a. Carbon nanotubes are grown all over the sample, using a standard CVD-growth recipe based on CH$_{4}$ feedstock gas and using Mo-Fe nanoparticles as catalyst \citep{Viennot2014c}. Depending on the catalyst distribution, several carbon nanotubes will be suspended between adjacent cantilevers. The large number of cantilevers is particularly useful to integrate multiple nanotubes on the same circuit chip. Prior to the transfer of a carbon nanotube to the circuit chip, pictures of the tips of all the cantilevers are taken using Scanning Electron Microscope (SEM) at low acceleration voltage (2 kV). While common belief is that SEM observation induces the deposition of hydrocarbon layer onto the nanotube \cite{Choi2016}, hence should introduce disorder in the electronic spectrum of the nanotube, we have found that it is still possible to obtain clean transport spectra using such a brief observation. One possible explanation for this low contamination of suspended carbon nanotubes by the e-beam exposure is the fact that most of the hydrocarbon is on the surface of the substrate, thus suspended nanotubes are much less affected by e-beam induced diffusion of hydrocarbon \cite{Wong2006}. This step is the first characterization allowing us to pre-select the isolated carbon nanotubes. Images of carbon nanotubes suspended in between adjacent cantilevers are shown in figure \ref{figure1}.a. Between the localization of carbon nanotubes and the stamping process, the cantilever chip is stored in a vacuum chamber (pressure: $5.10^{-7}$ mbar), to minimize the exposition time of nanotubes to ambient atmosphere down to approximately 10 min. 

On the circuit chip, the niobium microwave cavity, and the circuit electrodes are fabricated using e-beam lithography as in previous works \cite{Viennot2015}. In addition, two trenches on both sides of the circuit are etched with reactive ion etching using SF$_6$ gas. This process allows us to etch 10 to 15 $\mu$m deep trenches, which are used to ensure a proper contacting of the nanotube to the circuit electrodes (see figure \ref{figure1}.c and \ref{figure1}.d).

\begin{figure*}[ht!]
\includegraphics[width=\linewidth]{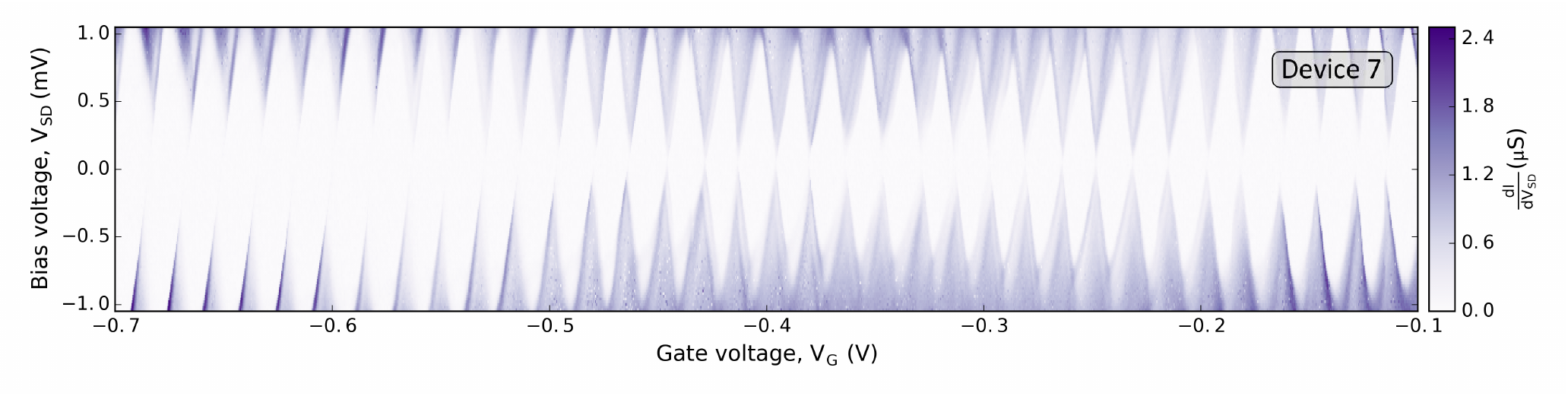}
\caption{\label{figure3} 
Differential conductance measured in device 7 versus bias voltage $V_{SD}$ and gate voltage $V_{G}$ at B = 0 T. The fact that the Coulomb diamonds do not close at $V_{SD} = 0$ V is a manifestation of the superconducting contacts (see Table \ref{table}). This 2D map is spanning over a large range of gate voltage in the Coulomb blockade regime (more than 30 Coulomb diamonds) and shows continuous evolution of the contacts transparency which indicates an electrostatic control of the coupling rates. From the contrast of the coulomb diamonds, one can notice that in the gate voltage region: -0.7 V $< V_G <$ -0.5 V, we have $\Gamma_{1} < \Gamma_{2}$, then for -0.4 V $< V_G <$ -0.3 V we have $\Gamma_{1} \simeq \Gamma_{2}$, and finally for -0.2 V $< V_G <$ -0.1 V, we have $\Gamma_{1} \simeq \Gamma_{2}$. This observation further emphasize the electrostatic control of the coupling rates.
}
\end{figure*}

The transfer of the carbon nanotube to the circuit chips is realized in a dedicated vacuum chamber with a base pressure of 5.10$^{-7} $ mbar. A schematic of this chamber is presented in figure \ref{figure2}.c. The circuit chip is placed on a rotary arm, so that it can be either in an horizontal position for the transfer of the nanotubes, or in a vertical position, facing an argon gun, in order to clean the surface of the contact electrodes, as well as removing unwanted nanotubes. The growth chip is fixed with a tilt angle of 45$^{\circ}$ with respect to the horizontal plane, to ensure a good visibility from the top view, while maintaining a small footprint of the circuit chip. The position of the growth chip is controlled by piezo-motors and micro-manipulators. There is also the possibility to isolate the growth chip from the main chamber, in order to protect the carbon nanotubes during the cleaning of the circuit chip with the argon gun, or during the replacement of the circuit chip.

The growth chip is lowered using the piezo-motor stage and the contact is detected by monitoring the current between the two external contact electrodes (at V$_{bias}$ = 0.5-1.5 V) and the two inner contacts, as depicted in figure \ref{figure1}.c. After the contact, the resistance of the circuit typically ranges between 10 M$\Omega$ to 100 G$\Omega$. Similarly to ref \citep{Collins2001b, Waissman2013c}, the two external sections of the nanotube are cut by driving a large current through it (typically between 10 $\mu$A and 20 $\mu$A). During this operation, all the other contacts and gates are set to a floating potential, to avoid accidental cutting of the central section. Two examples of I-V curves corresponding to the cutting of the nanotube are shown in inset of figure \ref{figure2}.b. The two curves are showing very similar features, such as the current and the bias voltage at which the nanotube is cut. This shows that the nanotube has almost identical properties at two different sections a few $\mu$m away from each other, attesting the cleanliness of the transferred nanotube.

\begin{figure}[ht!]
\includegraphics[width=\linewidth]{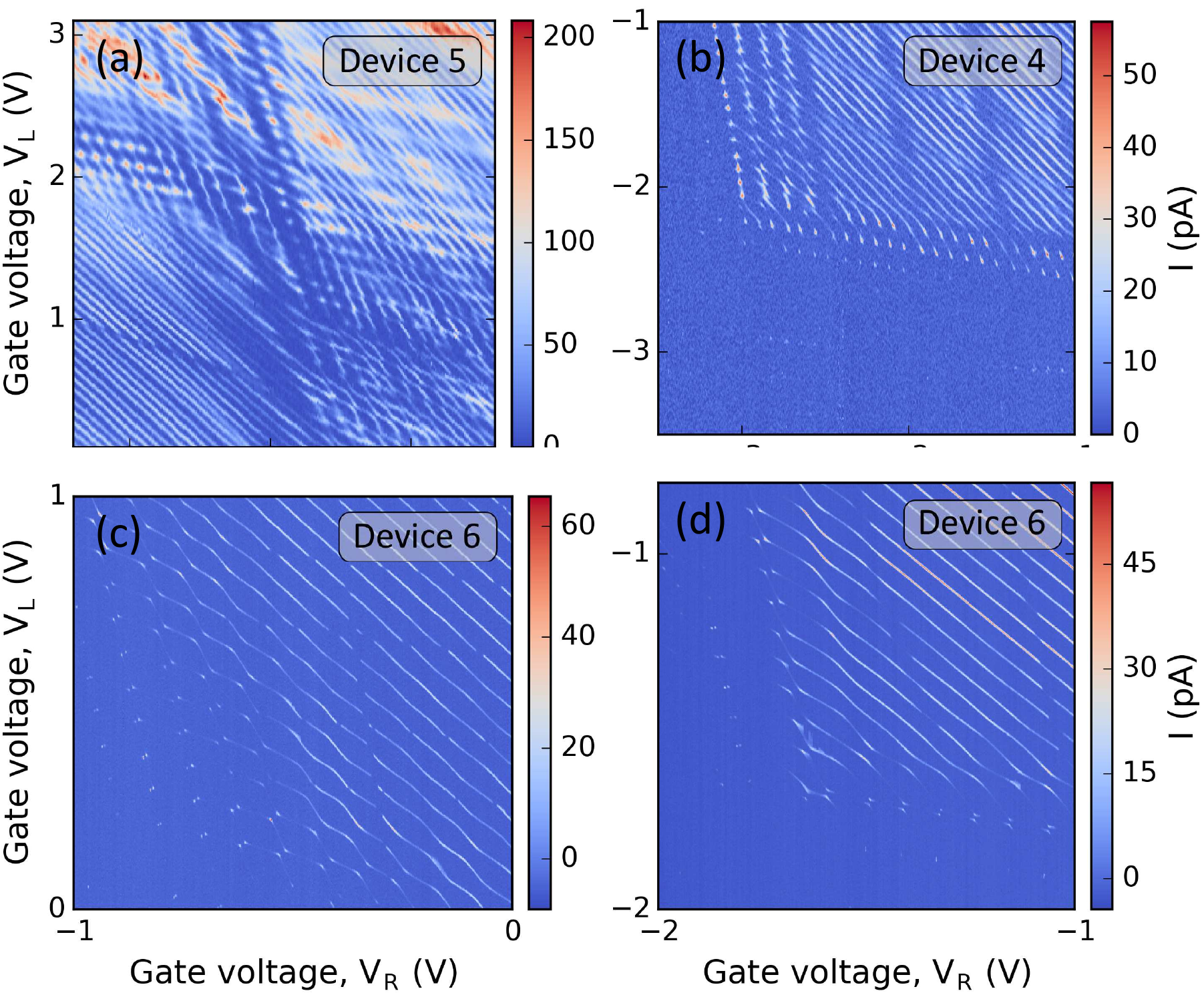}
\caption{\label{figure4} 
Double quantum dot stability diagram measured on three devices. 
\textbf{a.} Device 5 displays a narrow-gap behavior, and one can see the four different charge distributions depending on whether the two dots are filled with electrons or holes. Carbon nanotubes \textbf{b.} (Device 4), and \textbf{d.} (device 6) are semiconducting, and only the electron-electron filling distribution is visible in the gate-gate plane. \textbf{c.} For other gate voltages parameters of device 6, we observed a smooth transition from a single dot behavior (parallel lines in top right region) to a double quantum dot behavior (anti-crossing in bottom left region).}
\end{figure}

\begin{table*}[ht!]
\caption{Extensive characterization of 8 samples
\label{table} }
\begin{tabular}{ c | c | c | c | c | c | c | c  c }
\hline
\hline
 &    Contact metal   &    Geometry    &    Quality factor    &    RT resistance    &    tunnel rates $\Gamma_{1} / 2 \pi$    &    tunnel rates $\Gamma_{2} / 2 \pi$ & Charging energy\\
 & & & & (M$\Omega$) & (GHz) & (GHz) & (meV) \\
\hline
Device 1 & Au & 3 gates & 7400 & 0.08 & 0.5 & 16.5 & 1\\
Device 2 & Au & 3 gates & 16000 & 7 & 0.3 & 253 & 7\\
Device 3 & PdNi(25nm) / Pd(4nm) & 5 gates & 4000 & 2 & -- & -- & --\\
Device 4 & PdNi(25nm) / Pd(4nm) & 5 gates & -- & 0.8  & 0.5 & 0.5 & 2\\
Device 5 & PdNi(25nm) / Pd(4nm) & 5 gates & -- & 0.23 & 0.5 & 67 & 1.5\\
Device 6 & PdNi(25nm) / Pd(4nm) & 5 gates & 4000 & 0.55 & -- & -- & 2.2 - 2.4\\
Device 7 & Nb(45nm) / Pd(10nm) & 3 gates & -- & 0.2 & 33.6 & 3.7e-4 & 1.0\\
Device 8 & Nb(45nm) / Pd(10nm) & 1 gate & -- & 0.25 - 1.0 & 7 & 80 & 2.4\\
\hline
\hline
\end{tabular}
\end{table*}

One advantage of this cutting procedure is that it acts as a local annealing of the interface between the contact metal and the nanotube, drastically lowering the contact resistance, which can reach values below 1 M$\Omega$ (see table \ref{table}). The cutting step also allows to distinguish between a single nanotube which display a single current drop (see figure \ref{figure2}.b), and a bundle of nanotubes which display multiple current drops. At this stage, it is possible to further improve the contact quality by driving a high current in the central section of the device, being careful not to exceed the cutting current. If the contacted nanotube turns out to be a bundle or display unwanted characteristics, we remove it using the Ar gun. To obtain a suspended nanotube over the gate array, we found that the spacing of the contact electrode should not exceed ten times the height difference between contact and gate electrodes.

As a last characterization of the nanotube at room temperature, we measure the gate dependence of the current in the central section, in order to differentiate between a small-gap, a semiconducting or a metallic nanotube. Finally, during the transfer of the circuit chip into the cryostat, all the electrodes are grounded and the sample is maintain under a nitrogen atmosphere or a low vacuum.

The results discussed in the following are based on 8 different devices with various geometries and contact metals. The measurements have been performed at a base temperature of 20 mK (250 mK for device 8). We present in figure \ref{figure3}, a single dot stability diagram in the Coulomb blockade regime. The continuous evolution of the contact transparency as the gate voltage is swept as well as the absence of charge jumps are indicative of the clean environment of the nanotube. 

The tunability of the circuit parameters has also been observed in double quantum dot circuits. Figure \ref{figure4} presents stability diagrams of three double quantum dots devices with different carbon nanotube electronic behaviors: small-gap in figure \ref{figure4}.a and semiconducting in figure \ref{figure4}.b and d. The transition from double quantum dot behavior to a large single dot behavior is also visible in figure \ref{figure4}.c, showing again the weak influence of disorder on the confinement potential, which is instead dominated by electrostatic gating. The control over the interdot coupling was also demonstrated via cavity transmission measurements, where the transition from the resonant to the dispersive regime were observed (see supplementary material).

Importantly, we also applied this technique to multiple contacting materials. Table \ref{table} gathers informations on the 8 samples investigated in this article, including the coupling rates obtain with various materials. Using this transfer technique on ferromagnetic PdNi(25nm)/Pd(4nm) contacts allowed us to induce local polarization of electronic spin states in the nanotube and to couple it to cavity photons \citep{Cubaynes2019}. This technique can also be applied to superconducting contacts. Indeed, we measured a transport superconducting gap of $\Delta = 0.7 - 0.8$ meV, along with large coupling rates (7 GHz and 80 GHz), using Nb(45nm)/Pd(10nm) contacts. This last results demonstrate that this fabrication technique is fully compatible with hybrid circuits.

 We believe that the cleanliness of our quantum dot circuit, and the compatibility with various contacting material is partially due to the fact that the transfer process is performed under vacuum, which result in a cleaner nanotube-metal interface.

Transfering a carbon nanotube on top of circuit electrodes have proven to be a very efficent technique to integrate clean, and suspended carbon nanotubes into complex circuit designs with the possibility to be part of a cQED platform. Here, we present a new version of this transfer technique using commercial cantilevers, and we show that this approach can also be adapted to hybrid circuits containing superconducting and ferromagnetic materials, hence enlarging the scope of this technique to new experiments \cite{Mergenthaler2019, Desjardins2019, Khosla2018, Qin2019, Stadler2016}.

\begin{acknowledgments}
 The devices have been made within the consortium Salle Blanche Paris Centre. The authors gratefully acknowledge help from José Palomo, Aurélie Pierret, and Michael Rosticher. This work is supported by the ERC Starting Grant “CirQys”, the ERC Proof of Concept grant “QUBE”, the ANR “FunTheme”, and the Quantera grant “SuperTop”.
\end{acknowledgments}

\section*{Data Availability}

The authors declare that the main data supporting the findings of this study are available within the article. Extra data is available from the corresponding authors upon reasonable request.

\bibliography{library}

\end{document}